\documentclass[aps,prl,twocolumn,showpacs,letterpaper,superscriptaddress]{revtex4-1}
\pdfoutput=1
\usepackage{graphicx}
\usepackage{epstopdf}
\usepackage{multirow}
\usepackage{amsmath}
\usepackage{color}
\usepackage{hyperref}
\usepackage{makecell}
\usepackage{mathtools}
\usepackage{tabularx}
\usepackage{multirow}
\hypersetup{
    colorlinks=true,
   linkcolor=blue,
    filecolor=blue,
		urlcolor=blue,
   citecolor=blue,
}
\usepackage[normalem]{ulem}

\begin{document}

\title{Oxygen passivation mediated tunability of trion and excitons in MoS$_2$}

\author{Pranjal Kumar Gogoi}
\email{phypkg@nus.edu.sg}
\affiliation{Department of Physics, Faculty of Science, National University of Singapore, Singapore 117542}
\affiliation{Singapore Synchrotron Light Source, National University of Singapore, 5 Research Link, Singapore 117603}

\author{Zhenliang Hu}
\affiliation{Department of Physics, Faculty of Science, National University of Singapore, Singapore 117542}

\author{Qixing Wang}
\affiliation{Department of Physics, Faculty of Science, National University of Singapore, Singapore 117542}

\author{Alexandra Carvalho}
\affiliation{Centre for Advanced 2D Materials and Graphene Research Centre, National University of Singapore, Singapore 117542}

\author{Daniel Schmidt}
\affiliation{Singapore Synchrotron Light Source, National University of Singapore, 5 Research Link, Singapore 117603}

\author{Xinmao Yin}
\affiliation{Department of Physics, Faculty of Science, National University of Singapore, Singapore 117542}

\author{Yung-Huang Chang}
\affiliation{Department of Electrophysics, National Chiao Tung University, Hsinchu 30010, Taiwan}

\author{Lain-Jong Li}
\affiliation{Physical Science and Engineering Division, King Abdullah University of Science and Technology (KAUST), Thuwal, Saudi Arabia, 23955}

\author{Chorng Haur Sow}
\affiliation{Department of Physics, Faculty of Science, National University of Singapore, Singapore 117542}
\affiliation{Centre for Advanced 2D Materials and Graphene Research Centre, National University of Singapore, Singapore 117542}

\author{A. H. Castro Neto}
\affiliation{Department of Physics, Faculty of Science, National University of Singapore, Singapore 117542}
\affiliation{Centre for Advanced 2D Materials and Graphene Research Centre, National University of Singapore, Singapore 117542}

\author{Mark B. H. Breese}
\affiliation{Department of Physics, Faculty of Science, National University of Singapore, Singapore 117542}
\affiliation{Singapore Synchrotron Light Source, National University of Singapore, 5 Research Link, Singapore 117603}

\author{Andrivo Rusydi}
\email{phyandri@nus.edu.sg}
\affiliation{Department of Physics, Faculty of Science, National University of Singapore, Singapore 117542}
\affiliation{Singapore Synchrotron Light Source, National University of Singapore, 5 Research Link, Singapore 117603}
\affiliation{NUSNNI-NanoCore, National University of Singapore, Singapore 117576}

\author{Andrew T. S. Wee}
\email{phyweets@nus.edu.sg}
\affiliation{Department of Physics, Faculty of Science, National University of Singapore, Singapore 117542}
\affiliation{Centre for Advanced 2D Materials and Graphene Research Centre, National University of Singapore, Singapore 117542}
\date{\today}
%
%
\begin{abstract}

Using wide spectral range \emph{in situ} spectroscopic ellipsometry with systematic ultra high vacuum annealing and \emph{in situ} exposure to  oxygen, we report the complex dielectric function of MoS$_2$ isolating the environmental effects and revealing the crucial role of unpassivated and passivated sulphur vacancies. The spectral weights of the A ($1.92$~eV) and B ($2.02$~eV) exciton peaks in the dielectric function reduce significantly upon annealing, accompanied by spectral weight transfer in a broad energy range. Interestingly, the original spectral weights are recovered upon controlled oxygen exposure.
This tunability of the excitonic effects is likely due to passivation and reemergence of the gap states in the bandstructure during oxygen adsorption and desorption, respectively, as indicated by \emph{ab initio} density functional theory calculation results. This work unravels and emphasizes the important role of adsorbed oxygen in the optical spectra and many-body interactions of  MoS$_2$.


 \end{abstract}

\maketitle

The  current vigorous  research efforts in the field of ultrathin two dimensional transition metal dichalcogenides (TMDCs) have paid rich dividends with results showing fundamental physical implications as well as potential for applications~\cite{li_LJ, wang2012, Chhowalla, Mak_review_2016}. One of the most prominent aspects of the electronic structure and optical properties of these materials is the presence of strong many-body interactions~\cite{Evans, Wilson}. These are manifested as strongly bound excitons and other types of exciton complexes such as trions ~\cite{MakNatM_2013} and biexcitons~\cite{You_Yumeng}. Among the TMDCs, molybdenum disulfide (MoS$_2$) has been one of the most widely studied. As the understanding and optimization of the optical properties are of prime importance for TMDCs in general and MoS$_2$ in particular, there have been considerable efforts to manipulate its optical properties~\cite{MakNatM_2013, Newaz, Liu_Bo, Lin_Y, Amani, Mouri, Tongay_NL, Li_Ziwei}. 
Due to its two dimensional nature,  properties of monolayer MoS$_2$ are influenced strongly by the ambient~\cite{Lin_Y, KC_Santosh, Qi_Long}. Typically monolayer MoS$_2$ prepared by mechanical exfoliation and chemical vapour deposition (CVD) is known to have considerable sulphur vacancies (up to $\sim$3.5 $\times$ 10$^{13}$ cm$^{-2}$)~ together  with other kinds of defects~\cite{Zhou_Wu, Hong_Jinhua, Nan}. These sulphur vacancies contribute to deep acceptor levels in the bandstruture~\cite{Zhou_Wu, Komsa_PRB2015}. 

Despite the fundamental nature of the problem of the role of the sulphur vacancies and oxygen adsorption on the optical properties of MoS$_2$, there has been no report of the complex dielectric function yet, measured in a controlled ultrahigh vacuum (UHV) conditions together with systematic annealing and oxygen exposure.   Moreover, an unambiguous understanding of the underlying mechanism of optical property variations due to oxygen adsorption as well as sulphur vacancies based on combined and comprehensive experimental and theoretical work is lacking until now. Previous studies on the role of these defects as well as oxygen adsorption on the optical properties were performed only in the context of photoluminescence (PL)~\cite{Tongay_NL, Nan}. These works claimed the mechanism of  PL modulation to be based on transfer of electrons from the oxygen molecule in the adsorption sites~\cite{Tongay_NL, Nan}. However, the mechanism of PL modulation upon oxygen adsorption is unclear as the charge transfer mechanism at best could only contribute partially to the observed PL modulation~\cite{Liu_Yuanyue, Shu_ACSApplMatInt}. Moreover, there are no reports where the role of exciton and trions are explored or elucidated in detail based on \emph{in situ} experiments in controlled UHV environments. 

In this work, fully \emph{in situ} spectroscopic ellipsometry (SE) is employed to determine the  complex dielectric function of large area monolayer MoS$_2$ as a function of annealing temperature and oxygen exposure duration. The spectral weights of the exciton and trion peaks reduce systematically upon UHV annealing at higher temperatures. 
Upon subsequent oxygen exposure at 300~K, the initial complex dielectric function is recovered, which indicates  that oxygen adsorption/desorption is the cause of these changes. This UHV annealing and subsequent oxygen exposure cycle can be used for reduction and recovery of the exciton and trion spectral weights in a repeatable way. Density functional theory (DFT) based bandstructure calculations indicate that the exciton and trion spectral weight modulation is intricately related to the passivation and appearance of the gap states due to sulphur vacancies.

Large area monolayer MoS$_2$ grown directly on sapphire by the CVD method has been used for this work~\cite{Zhang_Xin, Suppl}. 
In-plane complex dielectric function is extracted for the energy range of 0.6-6.0 eV using a rotating analyzer spectroscopic ellipsometer with a compensator (Woollam V-VASE)~\cite{Fujiwara, Suppl}.  This wide spectral range allows for the determination of important optical transition peaks as well as corresponding spectral weights involving  parts of the valence and conduction bands further from the direct band edge. Spectroscopic ellipsometry  measurements are performed at 300~K and inside  a Janis cryostat at UHV conditions with  base pressure in the  10$^{-8}$ torr regime unless otherwise specified.

\begin{figure}
\includegraphics[width = 5.5in, clip, trim=40 40 0 0]{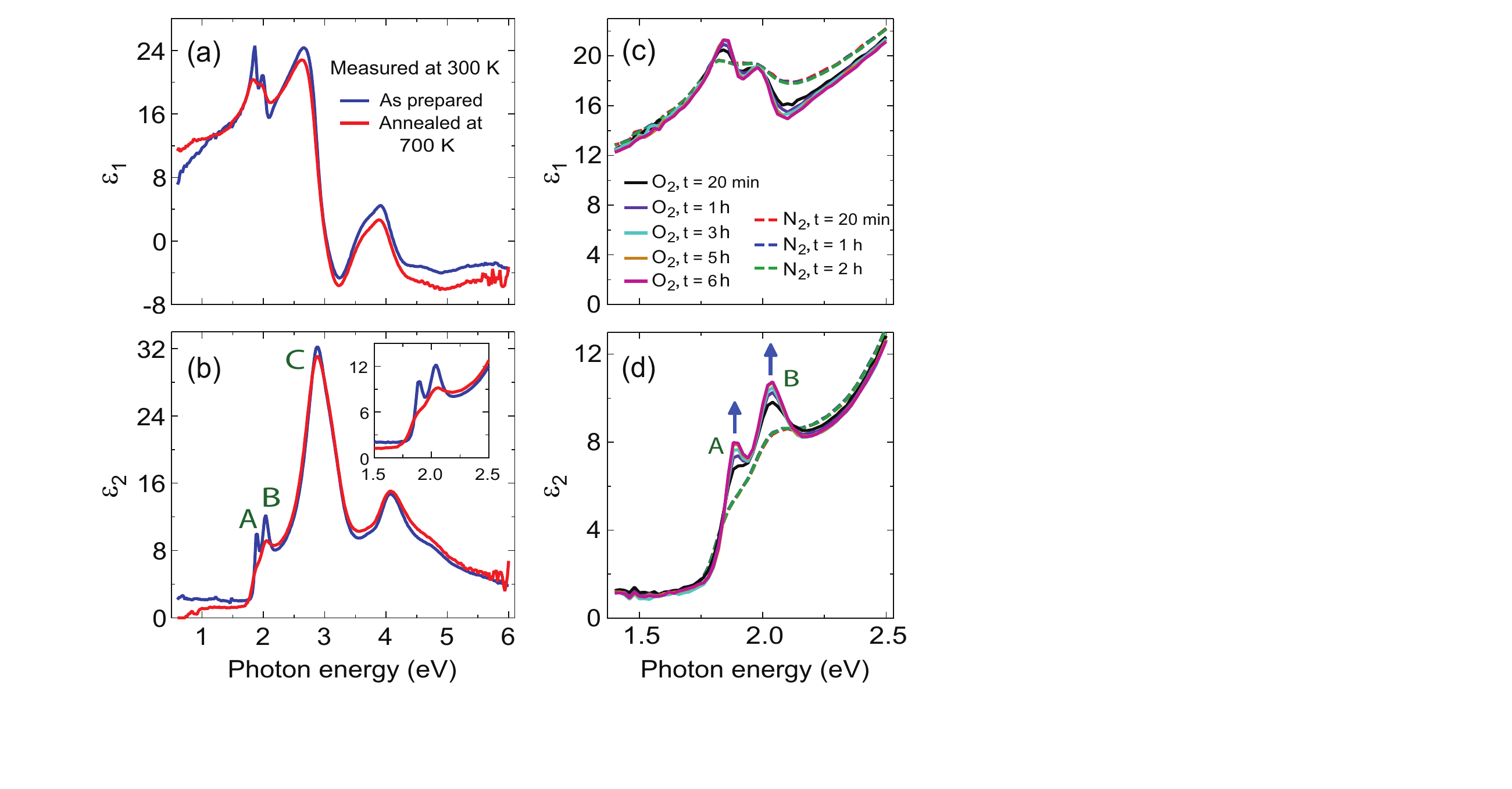}
\caption{Optical spectra of monolayer MoS$_2$ measured at 300~K. The comparison of  $\varepsilon_1$ ($\varepsilon_2$) measured before and after annealing in UHV at 700~K is shown in (a) ((b)). Evolution of the excitonic peaks A and B are shown in (c) and (d) when exposed to nitrogen and oxygen after annealing at 700~K. The direction of evolution of the peak structures (with time) is indicated by the arrows. Note that results shown in  (a) and (b) are from UHV measurements, while  (c) and (d) are from measurement where the sample chamber is flushed with the respective gases.}\label{fig1}
\end{figure}

We start our discussion for the sample that is systematically subjected to annealing treatments in the range of 300-700 K in UHV conditions at 50 K stepwise intervals, holding at a particular temperature for an hour. Spectroscopic ellipsometry spectra are acquired after cooling to 300 K. In  Figs.~\ref{fig1}(a) and \ref{fig1}(b), a representative complex dielectric function ($\varepsilon = \varepsilon_1 + i\varepsilon_2$) is shown  both for the pristine case  as well as after annealing at 700~K. The complex dielectric function spectra for the pristine MoS$_2$ show characteristic features dominated by the sharp peaks A, B at 1.92 and 2.02~eV, and the broad peak C at 2.80~eV as can be seen in the $\varepsilon_2$ (Fig.~\ref{fig1}(b)). The peaks A and B have been interpreted as bound excitons while the peak C as resonant exciton~\cite{Ramasubramanium, Berkelbach, QiuPRL, LiNguyenPRB, LiPRB, Zhang_Changjian, QiuPRL_Erratum}. The bound excitonic peaks A and B are associated with the direct transitions from the spin-orbit split valence band to the conduction bands at the $K$ (or $K^\prime$) point in the Brillouin zone. On the other hand, the resonant excitonic peak C  is associated with the transitions in the region around the $\Gamma$ point.    Interestingly, after annealing in UHV, the intensities and sharpness of the A and B excitons are reduced strongly accompanied by spectral weight transfer over a broad energy range (up to 6 eV, which is the energy limit of our measurements).

We next perform \emph{in situ} SE experiments under different gas exposures. After annealing at 700~K and cooling down to 300~K, the sample is exposed to nitrogen gas. As shown in Figs.~\ref{fig1}(c) and \ref{fig1}(d), the complex dielectric function does not change even after a few hours or longer. Interestingly, upon oxygen exposure at 300~K, the A and B excitonic peaks show gradual recovery to their original intensities and sharpness. It is observed that the peak intensities reach above 90\% of its original intensity after 10~hours of oxygen exposure.  Generally, CVD grown MoS$_2$ samples are $n$-doped and this $n$-doping is partially compensated by charge transfer from  physisorbed oxygen molecules~\cite{Tongay_NL, Nan}. However, as explained later, chemisorption of oxygen molecules in the sulphur vacancy sites further plays a crucial role in the passivation of the gap states which results in an almost clean gap. Upon progressive heat treatment in UHV, the oxygen is desorbed making the sample more $n$-doped (due to desorption of physisorbed oxygen) and also with gap states (due to desorption of chemisorbed oxygen). Upon exposure to oxygen, adsorption occurs again and this desorption and adsorption cycle can be used as a reversible method to manipulate the exciton and trion spectral weights (Figs.~\ref{fig1}(c), \ref{fig1}(d)). Independent measurements on the substrate alone show negligible changes in the substrate dielectric function with the annealing treatment or oxygen exposure, which eliminates the role of the substrate in the observations. 

\begin{figure}
\includegraphics[width = 3.2in, clip, trim=0 20 20 0 ]{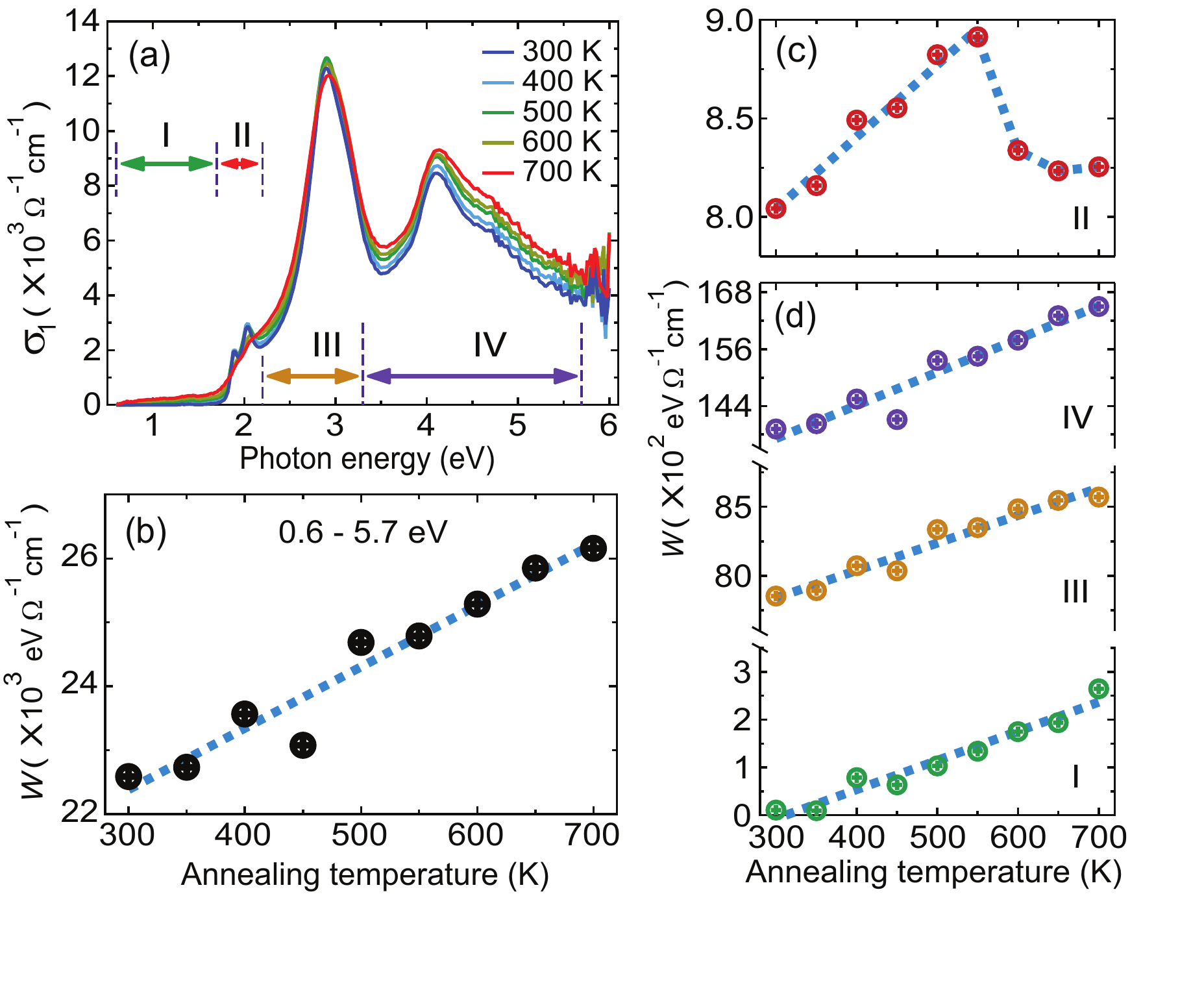}
\caption{ Real part of the optical conductivity and spectral weight for different annealing temperatures and measured at 300 K after cooling down. Real part of the optical conductivity for different annealing temperatures are shown in (a). The spectral weight for the full range (0.6-5.7~eV) and the region II (1.7-2.2~eV) are plotted in  (b) and (c), respectively. The spectral weight for regions I (0.6-1.7~eV) , III (2.2-3.3~eV)  and IV (3.3-5.7~eV) are shown in (d). In (b), (c) and (d) the dashed lines are guides to the eye.} \label{fig2}
\end{figure}

The real part of the optical conductivity $\sigma_1$, which is used to analyze spectral weight transfer, is plotted in Fig.~\ref{fig2}(a) for representative annealing temperatures in steps of  100~K. The $\sigma_1$ plots clearly show the evolution of the peaks and structures over the whole spectral range as a function of annealing temperature. To analyze and understand the evolution of $\sigma_1$, the full energy range is divided into four regions which individually include important features.  Regions I, II, III and IV are from 0.6-1.7, 1.7-2.2, 2.2-3.3, and 3.3-5.7~eV, respectively. The partial spectral weight can be calculated from $\sigma_1$ for each region using  the relationship, $W = \frac{{\pi}N_{\rm eff}e^2}{2m_{\rm e}} = \int_{E_1}^{E_2} \sigma_1(E) dE$, where $N_{\rm eff}$ is the electron density taking part in optical transitions in that energy interval (from $E_1$ to $E_2$), $e$ is the elementary charge, and $m_{\rm e}$ is the effective mass of the electron. As shown in  Fig.~\ref{fig2}(b) the total spectral weight for the full range (0.6-5.7~eV) increases almost linearly with the increase of annealing temperature. Note that the highest energy is taken as 5.7~eV instead of 6.0~eV to avoid experimental data noise.  Similarly, an overall linear trend is observed   for the regions I, III and IV (Fig.~\ref{fig2}(d)). However, the spectral weight for region II, which contains the A and B excitonic peaks, shows an anomalous decrease beyond $\sim$550~K (Fig.~\ref{fig2}(c)). This implies that although the overall effective electron density is increasing for the whole spectral range, the number of charges taking part in the optical transitions are in fact decreasing in the spectral range containing the A and B excitons. 

\begin{figure}
\includegraphics[width = 3.2in, clip, trim=30 30 20 0 ]{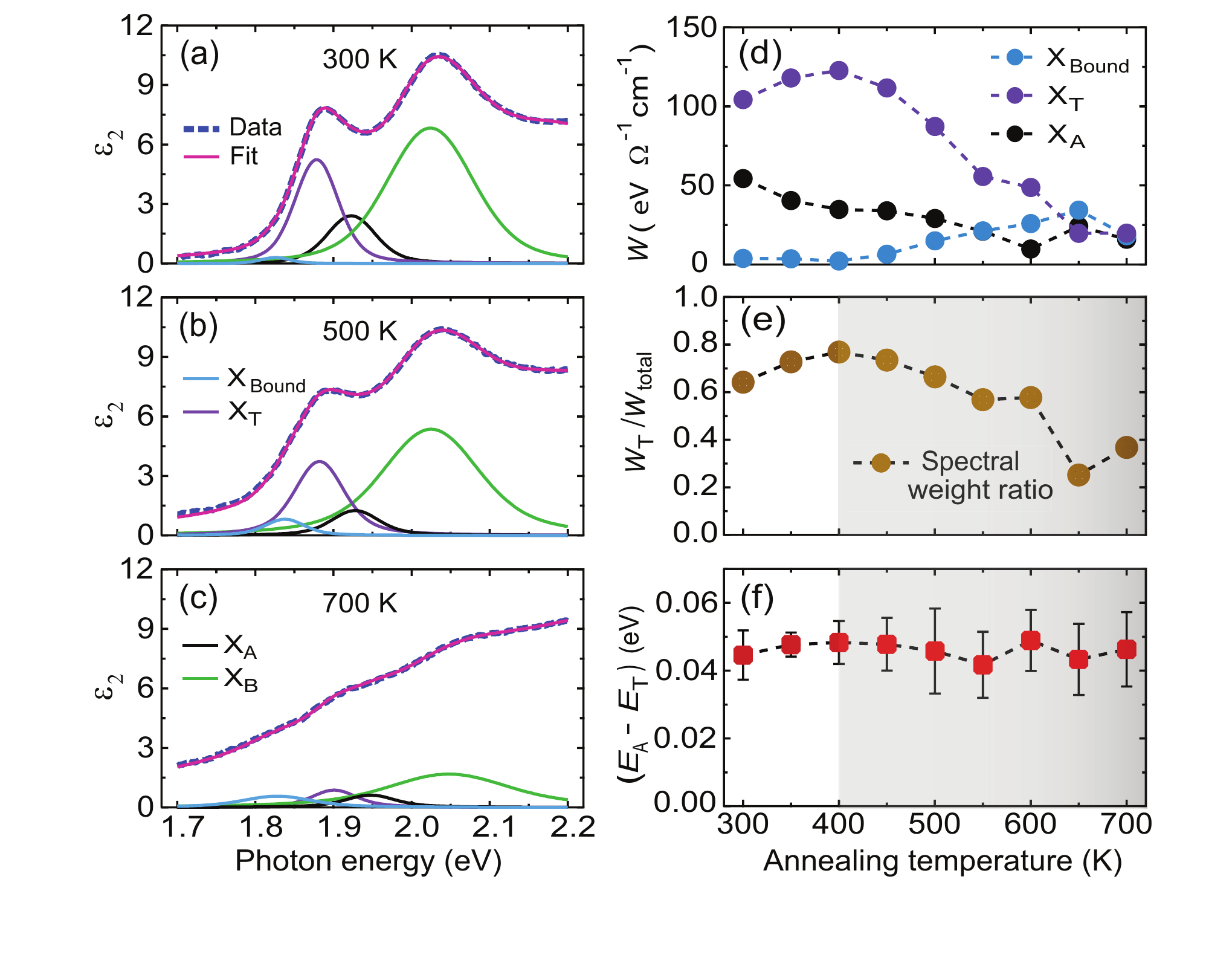}
\caption{ Contributions of excitons and trion to the optical spectra of MoS$_2$. Data and fit of $\varepsilon_2$ using Lorentzian-Gaussian lineshapes for X$_{\rm Bound}$, X$_{\rm T}$, X$_{\rm A}$, and X$_{\rm B}$  are shown in (a), (b), and (c) for annealing temperatures of 300, 500,  and 700~K, respectively.  Spectral weights of X$_{\rm Bound}$, X$_{\rm T}$, and X$_{\rm A}$ are depicted in (d). The ratio of the spectral weight of  X$_{\rm T}$ to X$_{\rm T}$, X$_{\rm A}$, and X$_{\rm Bound}$ together against annealing temperature are shown in (e). Difference in the energy positions for X$_{\rm A}$ and X$_{\rm T}$ with respect to annealing temperature is plotted in (f). In (d), (e), and (f) the dashed lines are guides to the eye. The shaded portions in (e) and (f) indicate regions where desorption of chemisorbed oxygen affects the respective quantities conspicuously.} \label{fig3}
\end{figure}

To better understand the origin of the individual components of the optical transition strengths, the complex dielectric function is extracted  from a finer grid (2~meV) of SE experimental data taken for the energy range 1.7-2.2~eV using wavelength-by-wavelength fit. The complex dielectric function is then modeled using  mixed Lorentzian-Gaussian lineshapes maintaining Kramers-Kronig consistency~\cite{Kim_C}. Five oscillators each with 30\% Lorentzian and 70\% Gaussian character give the best fit for the complex dielectric function at 300~K (with no annealing). For the intermediate temperatures, best fits are obtained when this ratio is linearly interpolated with the contributions being 50\% each for the highest annealing temperatures of 700~K~\cite{Suppl}. The substantial Gaussian contribution to the broadening indicates that the perturbation to the local potential from defects and impurities plays a dominant part in the observed inhomogenous lineshape broadening~\cite{MakNatM_2013, Sim, Chow_PK, Moody}. 
Representative fits of $\varepsilon_2$ are shown for the temperatures of 300, 500 and 700~K in Figs.~\ref{fig3}(a), (b) and (c), respectively. The fit details and results for all the annealing temperatures including  $\varepsilon_1$ can be found in the supplemental material~\cite{Suppl}. From the fit results for 300~K, the lineshapes with peak positions at 1.878, 1.923 and 2.024~eV could be identified as the trion (X$_{\rm T}$), A exciton (X$_{\rm A}$) and B exciton (X$_{\rm B}$)~\cite{MakNatM_2013, Sim, Mouri, Lin_Y, Chow_PK, Zhang_Changjian}. Interestingly, there is a small peak at 1.826~eV, which we attribute to the so called defect-bound exciton (X$_{\rm Bound}$)~\cite{Chow_PK}. The last broad lineshape at 2.191~eV mostly accounts for the large background and could have substantial contribution from the first higher order exciton corresponding to X$_{\rm A}$~\cite{QiuPRL, Hill}. 

The spectral weights calculated using individual lineshapes for X$_{\rm Bound}$, X$_{\rm T}$, and X$_{\rm A}$ are plotted in Fig.~\ref{fig3}(d). One of the important observations is that the X$_{\rm T}$ spectral weight is much larger than the X$_{\rm A}$ spectral weight, which is consistent with  the intrinsic $n$-doped nature of the sample~\cite{MakNatM_2013, Liu_Bo, Mouri}. Upon  annealing the sample becomes increasingly $n$-doped and the X$_{\rm T}$ spectral weight  increases slowly until $\sim$400~K. On the other hand, the spectral weight of X$_{\rm A}$ starts to decrease gradually upon annealing. 
It is to be noted that the total spectral weight of X$_{\rm T}$, X$_{\rm A}$, and X$_{\rm Bound}$  is dominated by X$_{\rm T}$ until 650~K, where X$_{\rm Bound}$ also becomes significant. Importantly, starting from 400~K the spectral weight decreases considerably for X$_{\rm T}$, which could  plausibly be attributed mainly to the annihilation of excitons due to charged defects states in the sulphur vacancy sites as explained later~\cite{Safarov}. The spectral weight of X$_{\rm Bound}$ starts to grow with annealing temperature after 400~K and becomes comparable in magnitude to that of X$_{\rm T}$ at $\sim$650~K. This trend is consistent with our conclusion that sulphur vacancies, which were previously occupied and passivated by oxygen, now after oxygen desorption, introduce gap states~\cite{Attaccalite}. The ratio of the spectral weight $W_{\rm T}/W_{\rm total}$, where $W_{\rm T}$ is the spectral weight of X$_{\rm T}$, and  $W_{\rm total}$ is the total spectral weight of X$_{\rm A}$, X$_{\rm T}$ and X$_{\rm Bound}$, is representative of the electron concentration (Fig.~\ref{fig3}(e)), higher value reflecting more $n$-doping~\cite{MakNatM_2013, Mouri, Lin_Y, Ross}. 

\begin{figure}
\includegraphics[width = 3.2in, clip, trim=0 0 0 0 ]{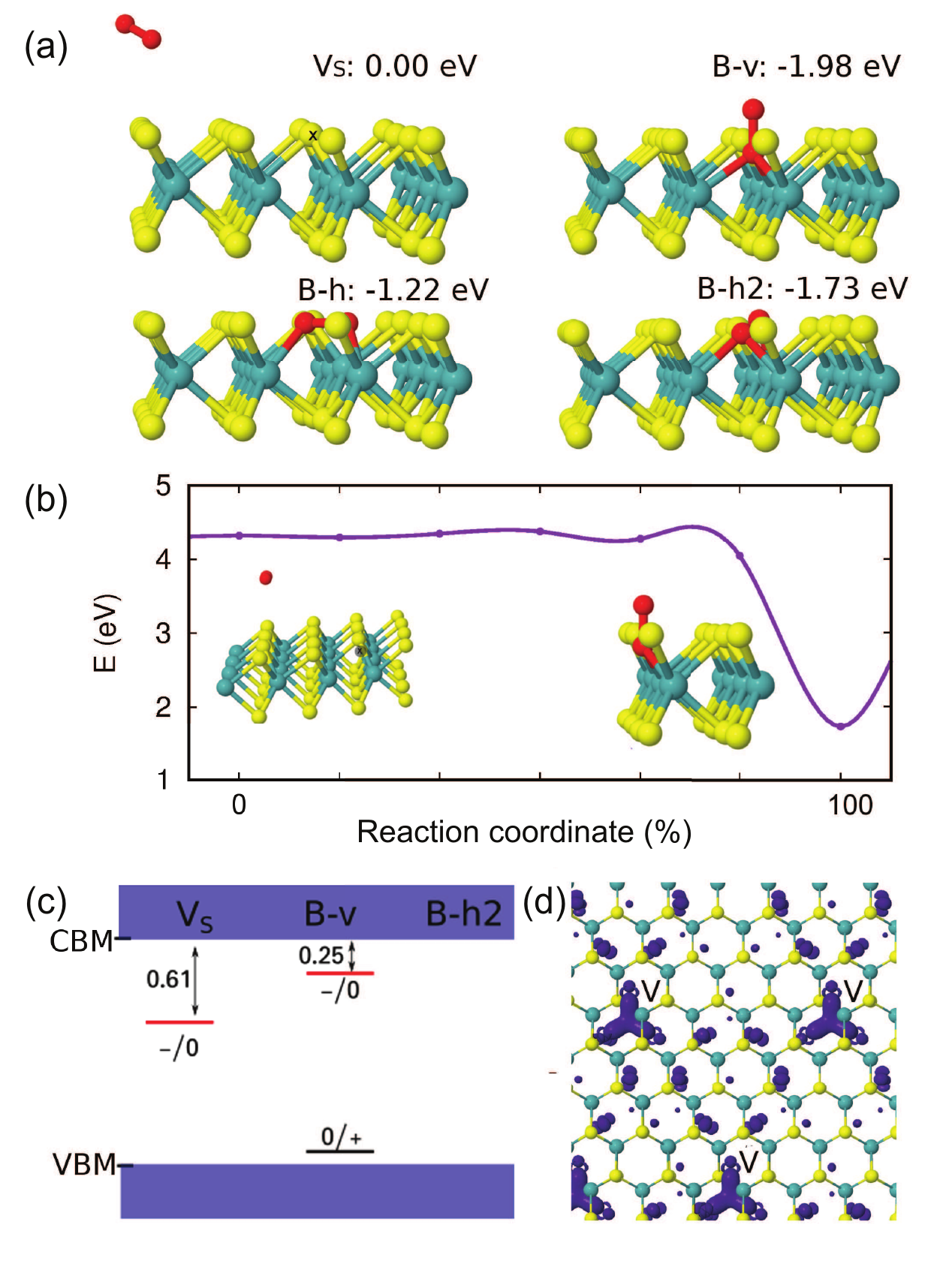}
\caption{Various configurations of oxygen adsorption with corresponding energies. (a) Sulphur, molybdenum and oxygen are represented by yellow, blue and red spheres, respectively. The sulphur vacancy site is represented by the cross. (b) Nudged elastic band calculation of the energy barrier for migration of an oxygen molecule towards a sulphur vacancy and respective trapping. The energy barrier, measured from the starting point, is 56 meV. The calculation was performed in the spin-averaged state. (c) Calculated ionization levels for relevant defects (all energies are in eV). VBM and CBM refer to the valence band maximum and conduction band minima, respectively. (d) Representation of the charge density of the trapped electrons at sulfur vacancies.}~\label{fig4}
\end{figure}

The trion binding energy can be estimated from the difference of the peak positions for X$_{\rm A}$ and X$_{\rm T}$. They are related by $E_{\rm A} - E_{\rm T} = E_{\rm {T_b}} + E_{\rm F}$, where $E_{\rm A}$, $E_{\rm T}$ are the exciton and trion peak positions, respectively; $E_{\rm{T_b}}$ is the trion binding energy and $E_{\rm F}$ is the Fermi level shift relative to the conduction band edge at the K point~\cite{MakNatM_2013}. 
This relationship can be used for a reasonable estimate of $E_{\rm T_b}$ for the range $\leq\sim$~400~K, assuming $E_{\rm F}$ to be small (Fig.~\ref{fig3}(f)). 
The estimate of $E_{\rm T_b}$ of $(45 \pm 7)$~meV for 300~K is in reasonable agreement with previous reports~\cite{MakNatM_2013, Lin_Y, Zhang_Changjian}.

To understand the role of sulphur vacancies and adsorbed oxygen on the electronic structure of MoS$_2$, \emph{ab intio} DFT based calculations are performed~\cite{Suppl}. The calculation results show that depending on the proximity and the orientation of the oxygen molecule there will be different energy configurations (Fig.~\ref{fig4}(a)). The electronic bandstructure of pristine MoS$_2$ is characterized by a clean gap as expected from a defect free material. However,  the presence of sulphur vacancies introduce deep acceptor levels, and new filled levels just above the valence band maximum of the pristine case~\cite{Suppl}\footnote{The acceptor levels of both V$_{\rm S}$ and B-v are deep because the energy necessary for them to accept an electron, creating free holes is much larger than the average thermal energy~(\emph{kT}).}. In the $n$-doped MoS$_2$ the Fermi level is close to the conduction band minimum and hence the deep acceptor level will be filled, and correspondingly the vacancy sites are expected to have negative charge (Fig.~\ref{fig4}(c) and~\ref{fig4}(d)). These charged localized vacancy sites should help in annihilation of the excitons whenever those are within a Bohr radius of excitons~\cite{Safarov}. A simple quantitative estimation of such vacancy site concentration can be obtained using the exciton Bohr radius, which is $\sim$1 nm for X$_{\rm A}$ for MoS$_2$~\cite{Zhang_Changjian, Sie}. Using this exciton Bohr radius the vacancy concentration expected for 50\% reduction in the exciton population is $\sim$1.6 $\times$ 10$^{13}$ cm$^{-2}$. This estimate is in the regime of  the observed  monosulphur vacancy concentration~\cite{Hong_Jinhua}, and thus suggests that the reduction of the spectral weight is possibly due to the annihilation of the excitons caused by charged defect sites. The deviation from stoichiometry reduces the density of states in the valence and conduction band leading to  further reduction of  the exciton and trion spectral weights. However, the drastic reduction of the exciton and trion spectral weights in comparison to other features indicates that the annihilation of excitons has the dominant contribution.

The physisorption of oxygen molecule in pristine MoS$_2$ has negligible effect on the bandstructure and it is characterized by electron transfer from MoS$_2$ to oxygen~\cite{Suppl, Tongay_NL}. Importantly, the energy barrier for chemisorption in a sulphur vacancy site is very low and this can happen at room temperature~\cite{Liu_Yuanyue, Suppl}. The chemisorption of the oxygen molecule in the vacancy site results in an energy gain of 1.98 eV (Fig.~\ref{fig4}(b)), and the acceptor level of the vacancy is occupied by the oxygen 2$p$ electrons. This process results in a recovery to the near clean gap state~\cite{Suppl}. 
The `horizontal' configuration is closer to the clean gap. However, the lowest energy state is characterized by the `vertical' orientation of the oxygen molecule. The crucial passivation of the gap states by these chemisorbed oxygen molecules leads to the disappearance of the local charged vacancy sites and hence the excitons and trion formation is not affected.  

In summary, using systematic \emph{in situ} spectroscopic ellipsometry with  UHV annealing we report the complex dielectric function of MoS$_2$ isolating the environmental effects and unveiling the role of sulphur vacancies.  With controlled oxygen exposure, we have further demonstrated tunability of absorption features associated with trion and excitons in MoS$_2$. This comprehensive experimental and theoretical work explains the mechanism of the modulation of the exciton and trion features upon oxygen adsorption and desorption. 
Overall, this study highlights the importance of adsorbed oxygen in the passivation of the gap states in MoS$_2$, its implications in the optical spectra of MoS$_2$, and the need for adequate consideration of such extrinsic factors in the  proper understanding of the relevant results.
 
This work is supported by ASTAR Pharos grant R-144-000-359-305 and SERC 1527000012, Singapore National Research Foundation under its Competitive Research Funding (NRF-CRP 8-2011-06 and No. R-398-000-087-281), MOE-AcRF Tier-2(MOE2010-T2-2-121), and FRC (R-144-000-368-112), and the Medium Sized Centre Programme and CRP award ``\textit{Novel 2D materials with tailored properties: Beyond graphene}'' (R-144-000-295-281). The first-principles calculations were carried out on the Centre for Advanced 2D Materials and Graphene Research Centre high-performance computing facilities.

%

\end{document}